\documentclass[prd,nofootinbib,preprintnumbers,
]{revtex4}
\usepackage{amssymb}
\usepackage{amsmath}
\usepackage{mathtools}
\usepackage[dvipsnames]{xcolor}
\usepackage{mathtools}
\usepackage{dsfont}
\usepackage[hyperfootnotes=false]{hyperref}

\def\be{\begin{equation}}
\def\ee{\end{equation}}
\def\ba{\begin{eqnarray}}
\def\ea{\end{eqnarray}}

\def\lb{\label}
\def\dfrac{\displaystyle\frac}

\def\e{{\rm e}}

\renewcommand*{\email}[1]{\footnote{Electronic address: \href{mailto:#1}{\nolinkurl{#1}} }}

\renewcommand{\theequation}{\arabic{section}.\arabic{equation}}

\newcommand{\debug}[1]{{\iffalse\color{ultramarine} \textbf{#1}\fi}}

\begin{document}
\title{\begin{flushright}\begin{small} LAPTH-029/24 \end{small} \end{flushright} \vspace{1cm}
Mass formulas for supergravity black holes with string singularities }
\author{Igor Bogush$^a$\email{igbogush@gmail.com}}
\author{G\'erard Cl\'ement$^b$\email{gclement@lapth.cnrs.fr}}
\author{Dmitri Gal'tsov$^{c,d}$\email{galtsov@phys.msu.ru}}
\affiliation{
${}^a$ Moldova State University, strada Alexei Mateevici 60, 2009, Chi\c{s}in\u{a}u, Moldova
\\
$^b$ LAPTh, Universit\'e Savoie Mont Blanc, CNRS, France
\\
$^c$ Faculty of Physics, Moscow State University, 119899, Moscow, Russia
\\
$^d$ Kazan Federal University, 420008 Kazan, Russia }

\begin{abstract}
We extend the derivation of mass formulas for stationary axisymmetric asymptotically locally flat solutions with string singularities on the polar axis to general supergravity actions containing vector and scalar fields. It is based on the rod structure of the solutions in Weyl coordinates and is applicable to black holes with Dirac and Misner strings.  The obtained formulas differ from the corresponding ones in Einstein-Maxwell theory only by summation over all independent electric charges.

 \end{abstract}

\maketitle
\section{Introduction}
Recently, interest has arisen in the construction of mass formulas and the first law \cite{Smarr:1972kt,Hawking:1974rv,Bekenstein:1972tm,Bardeen:1973gs,Carter:1973rla} for black holes in the presence of conical singularities and Misner strings, which include systems of black holes with struts connecting them, as well as solutions with the Newman-Unti-Tamburino (NUT) parameter and the acceleration parameter \cite{Herdeiro:2010aq,Appels:2016uha,Clement:2017otx,Appels:2017xoe,Anabalon:2018qfv,Clement:2019ghi,Krtous:2019fpo, Hennigar:2019ive,Durka:2019ajz,Wu:2019pzr,Chen:2019uhp,BallonBordo:2020mcs,Gregory:2020mmi,Clement:2021ukb,Wu:2022rmx,Wu:2022mlz,Yang:2023hll}. The interest in NUTty solutions increased recently after it was shown that the suspected violation of chronology around Misner strings interpreted in Bonnor's spirit as physical distributional singularities may not be catastrophic \cite{Clement:2015cxa}. In the  works cited above the solutions of Einstein's vacuum equations or Einstein-Maxwell equations were mainly considered. The most general approach applicable to such systems suggested in Refs. \cite{Clement:2017otx,Clement:2019ghi} consists in the combination of the Komar-Tomimatsu method \cite{Komar:1958wp,Tomimatsu:1983qc,Tomimatsu:1984pw} of calculating the mass and angular momentum of individual black holes in the double-Kerr solution with the general theory of {\em rod structure} of stationary axisymmetric solutions developed in Refs. \cite{Harmark:2004rm,Emparan:2001wk}. Tomimatsu suggested the evaluation of Komar integrals for black hole masses and angular momenta in Weyl-Papapetrou coordinates in which the horizons look like infinitely thin cylinders along the polar axis. These integrals can be expressed through the Ernst potentials providing a three-dimensional sigma-model description of vacuum and electrovacuum configurations with two commuting Killing vector fields. A corrected version of the Tomimatsu approach was applied to stationary axisymmetric electrovac solutions in terms of the general theory of rod structure in Refs. \cite{Clement:2017otx,Clement:2019ghi}.
The resulting theory treats black hole horizons and Misner strings on an equal footing and also may account for acceleration horizons 
.
In the absence of Misner strings, it  was recently shown that such an approach also yields, through a limiting process, consistent mass formulas for aligned systems of black holes and Dirac strings in electrovacuum theory \cite{Clement:2023xvq}.

The purpose of the present paper is to generalize this method to extended supergravity theories in four dimensions involving the scalar fields.
Essential for this is the representation of solutions in terms of the potentials of three-dimensional sigma-models. This has been shown to work in the case of Einstein-Maxwell-Dilaton-Axion (EMDA) gravity, which is a truncated version of ${\cal N}=4$ supergravity \cite{Galtsov:2021iqx}. The sigma-model representation for more general supergravity actions has been considered in a number of papers including Refs. \cite{Breitenlohner:1987dg,Breitenlohner:1998cv,Clement:2008qx,Chow:2014cca}.
The corresponding Smarr-type mass formulas for 
supergravity black holes without singularities on the polar axis have been discussed, in particular,  in \cite{Gibbons:1982ih,Kallosh:1994ba,Gibbons:1996af,Chow:2014cca}. These mass formulas contain symmetric contributions from electric and magnetic charges, whose variations are multiplied by electric and magnetic potentials of the horizon. While this is correct for one-center solutions, the difference between electric and magnetic charges becomes manifest for multi-center solutions. Namely, magnetic potentials can not be uniquely prescribed to each constituent black hole, but depend on the neighbour's contribution. Magnetic contributions are associated with the Dirac strings connecting neighboring centers \cite{Clement:2023xvq}.

These features are naturally incorporated in the rod formalism, which treats black hole horizons and interconnecting Misner strings on an equal footing \cite{Clement:2017otx,Clement:2019ghi}. For each rod, one can derive a Smarr-type relation between the mass, angular momentum, electric charge, and surface area.
Then the total mass, angular momentum, and electric charge of the solution are expressed as the sum of the contributions of different rods. The resulting global quantities are not symmetric with respect to the pairs of electric and magnetic charges and the ADM mass and NUT parameters, respectively. 
Note that the separate contribution of the Misner string to the entropy of black holes endowed with NUT parameters was frequently discussed at the end of 90-ies \cite{Hunter:1998qe,Hawking:1998jf,Carlip:1999cy,Mann:1999pc}, suggesting an entropic interpretation of the area term in the mass formulas. In our opinion, a mechanical interpretation of this term is more appropriate.

\section{The setup}
We start with the action introduced by Breitenlohner, Maison and Gibbons \cite{Breitenlohner:1987dg,Breitenlohner:1998cv} for generic four-dimensional supergravity involving $n_s$ scalar moduli $\Psi^A$, $A=1,\ldots,n_s$ (dilatons and axions) and $n_v$ abelian vector fields $F^I=dA^I$, $I=1,\ldots,n_v$ (in slightly different notation, see Appendix \ref{app:sigma}):
\begin{equation} \label{eq:general_action}
 S =
 \int d^4x\left[
    \left(
          R
        - \frac{1}{2} f_{AB} \partial_\mu\Psi^A\partial^\mu\Psi^B
        - \frac{1}{2} K_{IJ} F^I_{\mu\nu} F^{J\mu\nu}
    \right) \sqrt{-g}
    - \frac{1}{2} H_{IJ} F^I_{\mu\nu} F^J_{\lambda\tau} \epsilon^{\mu\nu\lambda\tau}
  \right].
\end{equation} The  scalar moduli parametrize a four-dimensional coset (e.g. $U(8)/E_{7(7)}$ for  ${\cal N}=8, D=4$ supergravity) with an associated target metric $f_{AB}$. Vector fields transform under the same global symmetry implemented by real symmetric matrices $K_{IJ}$, $H_{IJ}$  depending on scalar fields $\Psi^A$ (summation over the repeated indices $I,\,J$ is understood). $R$ is the scalar curvature and the Levi-Civita tensor density is defined with $\epsilon_{0123}=1$. Also, we assume that $K_{IJ}$ is a positive definite matrix.

The corresponding equations of motion consist of the Einstein equations
\begin{equation}\lb{Ric}
    R_{\mu\nu} =
      \frac{1}{2} f_{AB} \Psi_{,\mu}^A\Psi_{,\nu}^B
    - K_{IJ} \left(
          F^I_{\mu\lambda}{F^{J\lambda}}_\nu
        + \frac{1}{4} g_{\mu\nu} F^I_{\alpha\beta}F^{J\alpha\beta}
    \right),
\end{equation}
scalar equations
\begin{equation}
    \frac{1}{\sqrt{-g}}
    \partial_\mu\left(
        \sqrt{-g} g^{\mu\nu} f_{AB}  \Psi^B_{,\nu}
    \right)
    =
    \frac{1}{2}\left(
          \frac{\partial f_{BC}}{\partial {\Psi^A}}
           \Psi^B_{,\mu} \Psi^{C,\mu}
        + \frac{\partial K_{IJ}}{\partial {\Psi^A}}
          F^I_{\mu\nu} F^{J\mu\nu}
        + \frac{\partial H_{IJ}}{\partial {\Psi^A}}
          F^I_{\mu\nu} F^J_{\lambda\tau} \frac{ \epsilon^{\mu\nu\lambda\tau}}{\sqrt{-g}}
    \right),
\end{equation}
Maxwell equations \begin{equation}\lb{Max}
    \partial_\mu\left( \sqrt{-g}G_I^{\mu\nu} \right) = 0,
\end{equation} written in terms of the induction tensors
\begin{equation}\lb{Gdef}
    G_I^{\mu\nu} =
      K_{IJ} F^{J\mu\nu}
    + H_{IJ} F^J_{\lambda\tau} \frac{\epsilon^{\mu\nu\lambda\tau}}{\sqrt{-g}},
\end{equation}
and Bianchi identities
\begin{equation}
    \partial_\mu\left( \epsilon^{\mu\nu\lambda\tau}F^I_{\lambda\tau} \right) = 0.
\end{equation}
Assuming stationarity of spacetime, implemented by a Killing vector $k=\partial_t$, we reduce the theory to three-dimensional sigma-model presenting the metric in Kaluza-Klein form
 \begin{equation}
    ds^2 = g_{\mu\nu}dx^\mu dx^\nu =
    - f(dt - \omega_idx^i)^2
    + f^{-1} h_{ij}dx^i dx^j,
\end{equation}
where the three-dimensional metric $h_{ij}$, the rotation  vector $\omega_i,\; (i, j=1, 2, 3)$ and the scale factor $f$ depend only on $x^i$. Derivation of the three-dimensional sigma model for the action (\ref{eq:general_action}) was  given
in
\cite{Breitenlohner:1987dg} and further detailed, in particular, in \cite{Breitenlohner:1998cv,Clement:2008qx,Chow:2014cca}. In our notation  (see  appendix \ref{app:sigma} for comparison with notation of
\cite{Breitenlohner:1987dg,Breitenlohner:1998cv}) basic steps are the following.
The spatial parts of the Bianchi identities and the modified Maxwell equations are solved by introduction of electric $v^I$ and magnetic $u_I$ potentials:
\begin{subequations}
\begin{align}\lb{vdef}
&F^I_{i0}=\partial_iv^I,\\\lb{udef}
&G_I^{ij}=\frac{f}{\sqrt{h}}\epsilon^{ijk}
\partial_k u_I,
\end{align}
\end{subequations}
while the mixed $R_0^i$ Einstein equations are solved by introducing the twist potential $\chi $ via   dualization equation
\begin{equation}\lb{eq:chi}
      f^2 h_{il} \frac{\epsilon^{ljk}}{\sqrt{h}} \partial_j \omega_k
    =u_I\partial_i v^I
    - v^I \partial_i u_I
    - \partial_i \chi.
\end{equation}
The remaining equations reduce to those of the three-dimensional gravitating sigma-model  for $n_\sigma=2+2n_v+n_s$ scalars $ X^M=f,\chi,\Psi_A,v^I,u_I, \;M=1,\ldots,n_\sigma$ and the three-dimensional metric $h_{ij}$:
\begin{equation}
S_3=\int \left[{\cal R}(h)-{\cal G}_{MN}(X)\,\partial_i X^M\,\partial_j X^N
\,h^{ij}\right]\sqrt{h}\,d^3x,
\end{equation}
where the target space metric ${\cal G}_{MN}$ can be presented as
\begin{align}\label{eq:sigma_model}
    &
    \mathcal{G}_{MN}dX^MdX^N=
      \frac{1}{2f^2} \left(
          df^2
        + \left(
              d\chi + v^I du_I - u_I dv^I
        \right)^2
    \right)
    + \frac{1}{2}f_{AB} d\Psi^A d\Psi^B
    \\\nonumber &
    - \frac{1}{f} \left(
          dv^I (K_{IJ} + 4 H_{IK} K^{KL} H_{LJ}) dv^J
        - 4 dv^I H_{IK} K^{KJ} du_J
        + du_I K^{IJ} du_J
    \right),
 \end{align}
and $K^{IJ}$ is the inverse matrix of $K_{IJ}$.

Also, the following identities for the Maxwell tensors will be useful:
\begin{subequations}
\begin{equation}
    F^{Ii0} = F^{Iij} \omega_j - h^{ij} F^I_{j0},\qquad
    F_{ij}^I = f^{-2}h_{ik}h_{jl} F^{Ikl} + F_{0i}^I\omega_j - F_{0j}^I\omega_i,
\end{equation}
\begin{equation}
    F_{ij}^I = \frac{h_{ik}h_{jl} \epsilon^{klm}}{f\sqrt{h}} K^{IJ}\left(
          \partial_m u_J
        - 2 H_{JK} \partial_mv^K
    \right)
    + \left(\partial_jv^I\omega_i - \partial_iv^I\omega_j \right).
\end{equation}
\end{subequations}

\section{Rod structure and Carter identities}
\label{sec:rods}
Further we assume axial symmetry under Killing translations along $ m=\partial_\phi$ and use the rod structure which was introduced to classify    $D$-dimensional vacuum spacetimes with $D-2$ commuting Killing vectors \cite{Harmark:2004rm}. This was shown \cite{Clement:2017otx,Clement:2019ghi} to be also applicable to Einstein-Maxwell four-dimensional solutions and helps to derive mass formulas and the first law of black hole mechanics in the presence of 
Dirac and Misner strings. Here we will show that the same is true for the  supergravity  black holes with scalar fields governed by the action (\ref{eq:general_action}).

Recall briefly basic definitions.
Let  $\gamma _{ab},\;x^a=t,\,\varphi$ be the two-dimensional Lorentzian metric of the subspace spanned by the Killing vectors.  Defining the    Weyl coordinates $\rho,\, z $  with
\begin{equation}             \label{rhodef}
 \rho = \sqrt{|\det \gamma |}\,,
\end{equation}
we present the spacetime metric as
 \begin{equation}               \label{Gzr}
 ds^2 = \gamma _{ab}(\rho, z) dx^a dx^b + e^{2\nu}(d\rho^2 +  dz^2)\,,
\end{equation}
where   $\nu$ is a function  of $(\rho, z)$. On the polar axis $\rho=0$ the matrix $\gamma_{ab}$ degenerates by virtue of the definition (\ref{rhodef}) and must have zero eigenvalues:
\begin{equation}\lb{dir}
 \gamma _{ab}(0, z)v_n^b = 0  \,.
\end{equation}
For quasiregular spacetimes, such that on the symmetry axis there are no strong curvature singularities, these eigenvalues are nondegenerate, except for a number of {\em turning points} $z_n$ splitting the polar axis
into finite or semi-infinite intervals (rods)
$(-\infty, z_1], [z_1, z_2], \dots, [z_N, +\infty)$, where
two semi-infinite rods are denoted by $n=\pm$, and the remaining finite rods by an index $n$ corresponding to the
left bound of the interval. For each rod $z\in [z_n, z_{n+1}]$,   a two-dimensional vector, {\em rod direction}, is a unique solution of (\ref{dir}).

By construction, the rod directions are associated with  Killing vectors in spacetime $\xi^\mu=v^0k^\mu+v^1 m^\mu$, for which  the submanifold $\rho=0$ will be the Killing horizon, i.e., $\xi^\mu\to l^\mu_n$ with $l^2=0$. Choosing normalization $v^0=1$ one finds that for each rod the angular velocity $\Omega_n$ and the associated surface gravity
\be\lb{kadef}
\varkappa_n= \left(|\xi_{\mu ;\nu} \xi^{\mu ;\nu}/2|\right)^{1/2}=\lim_{\rho\to 0} \left(- \rho^{-2}e^{-2\nu}\gamma_{ab}v^a_n v^b_n  \right)^{1/2},
\ee
are both constant on the respective rod (i.e., $z$-independent). The constancy of the surface gravity can be proven under the assumption of dominant energy condition \cite{Bardeen:1973gs}.

Another important property in the electrovacuum case is the constancy of the electrostatic potential on the horizon in the corotating frame. We have to generalize Carter's proof of this property to supergravity actions with multiple vector and scalar fields and for spacelike rods representing 
Misner strings.

One starts with Carter's identity
\be
    \label{eq:carter}
    R_{\mu\nu} l_n^\mu l_n^\nu=0
\ee
valid for generators  $l_n^\mu$ of the Killing horizon. Eq. (\ref{eq:carter}) can be obtained from Raychaudhuri identities following Ref. \cite{carter}. Using (\ref{Ric}) one can see that scalar fields do not contribute, so one is left with
\begin{equation}\label{eq:energy_momentum_l}
    K_{IJ}\left(
          F^I_{\mu\lambda}{F^{J\lambda}}_\nu
        + \frac{1}{4} g_{\mu\nu} F^I_{\alpha\beta}F^{J\alpha\beta}
    \right) l^\mu l^\nu = 0.
\end{equation}
Introducing  electric and  magnetic fields in  the co-rotating frame,
\begin{equation}\label{eq:corotating_eb}
    E^I_\mu = F_{\mu\nu}^I l^\nu,\qquad
    B^I_\mu = \tilde{F}_{\mu\nu}^I l^\nu,\qquad
\end{equation}
this condition can be rewritten as \cite{carter}
\begin{equation}\label{eq:energy_momentum_l2}
        K_{IJ}\left( E_\mu^I E^{\mu J} + B_\mu^I B^{\mu J} \right) = 0.
\end{equation}
Note that the left hand sides of Eqs. (\ref{eq:energy_momentum_l}) and (\ref{eq:energy_momentum_l2}) are identical to each other up to some constant factor in the entire spacetime, where $l^\mu$ is not necessarily null. Since $K_{IJ}$ is a positive definite matrix, this can be fair only if $\tilde{E}^J_\mu = U^J_{I} E_\mu^I$ and $\tilde{B}^J_\mu = U^J_{I} B_\mu^I$ are null, where $U^J_{I}$ diagonalizes the matrix $K_{IJ}$. Following from the antisymmetric form of $F_{\mu\nu}$ and $\tilde{F}_{\mu\nu}$, one can notice that $\tilde{E}_\mu^I l^\mu_n = 0$, $\tilde{B}_\mu^I l^\mu_n = 0$, thus vectors $\tilde{E}_\mu^I$, $\tilde{B}_\mu^I$ are orthogonal to $l^\mu_n$. The only null direction orthogonal to $l_n^\mu$ at the Killing horizon is $l_n^\mu$ itself, so vectors $\tilde{E}_\mu^I$, $\tilde{B}_\mu^I$ have $l_n^\mu$ direction, i.e., all of them have the same direction. Then, any linear combination has direction $l_n^\mu$, including the initial vectors $E_\mu^I$ and $B_\mu^I$. As a result, the following commutators are zero
\begin{equation}
    {E^I}_{[\mu} l_{\nu]} = 0,\qquad
    {B^I}_{[\mu} l_{\nu]} = 0.
\end{equation}
Particularly, choosing indices $\mu=i$, $\nu=t$, we find that ${E^I}_{i} = 0$, which can be rewritten using Eq. (\ref{eq:corotating_eb}) in the following form
\begin{align} \label{eq:electric_horizon}
    0
    = {E^I}_i
    = F_{i\nu}^I l^\nu
    = F_{it}^I l^t + F_{i\varphi}^I l^\varphi
    =
    \partial_i A^I_{t} + \Omega_n \partial_i A^I_{\varphi},
\end{align}
and since $\Omega_n$ is constant, we have constancy of the quantity $A^I_{t} + \Omega_n A^I_{\varphi}$ at the Killing horizon. In this proof  important assumptions were positive definiteness and diagonalizability of the matrix $K_{IJ}$. Similarly, in the magnetic sector, we have
\begin{align}\label{eq:magnetic_horizon}
    {B^I}_{i} = 0
    \quad\Rightarrow\quad
    \epsilon_{tij\varphi} \left(F^{Ij\varphi} - F^{I jt} \Omega_n\right) = 0
    \quad\Rightarrow\quad
    \left(F^{I jt} - \omega_n F^{Ij\varphi}\right)\sqrt{|g|} = 0,
\end{align}
where $\omega_n=1/\Omega_n$. Combining Eqs. (\ref{eq:electric_horizon}) and (\ref{eq:magnetic_horizon}), we obtain the following useful relation valid at the Killing horizon $l_n^\mu {l_n}_\mu =0$:
\begin{align} \lb{G0}
    \left(G_I^{\rho t}-\omega G_I^{\rho \varphi} \right)\sqrt{|g|} =
      K_{IJ}\left(F^{J\rho t}-\omega F^{J\rho \varphi} \right)\sqrt{|g|}
    - 2 H_{IJ}\epsilon^{t\rho z \varphi}\left( F^J_{z \varphi} + \omega F^J_{zt} \right)
  =0.
\end{align}

Denoting the scalar electric potential in the co-rotating frame
\be\label{elpot}
    \Phi^I = -(A^I_{t} + \Omega_n, A^I_{\varphi}),
\ee
we find out that it is constant along the given rod (i.e., at the Killing horizon) as far as $\Omega_n$ is constant, independently of whether the Killing vector is timelike or spacelike. The quantity $\Phi^I$ has a meaning of the scalar electric potential in the co-rotating frame corresponding to the one-form $A^I_\mu$.

Timelike rods of finite length correspond to black hole horizons, and infinite timelike rods describe acceleration horizons.
The angular velocities and surface gravities of horizon rods have the interpretation of angular velocities and Hawking temperatures. As we will see, the surface gravity rather has a mechanical interpretation for spacelike rods representing defects such 
Misner strings in the presence of NUTs.

It is believed (though rigorously proved only in vacuum gravity) that the rod structure uniquely fixes stationary axisymmetric asymptotically flat solutions. Thus, using Komar integrals over the cylinders surrounding rods one can find individual masses, angular momenta, and electric charges for each black hole and each defect. Asymptotic values of the same quantities can be then evaluated using Gauss-Ostrogradsky theorem including matter field contributions in the bulk. The corresponding balance equations were earlier derived in the electrovacuum  and dilaton axion-gravity and interpreted as Smarr mass formulas and the differential first law. An essential feature in the latter case was that dilaton and axion fields do not contribute to the bulk integrals for stationary axisymmetric configurations. We will see that the same property holds for generic extended supergravity action too, ensuring the possibility to generalize Smarr formulas for multicomponent solutions to supergravities.


\section{Smarr relations for generic rods}\label{sec:komar}
Following Ref. \cite{Clement:2019ghi}, we start with the Komar definition \cite{Komar:1958wp} of the asymptotic mass and angular momentum:
  \be
M_\infty = \frac1{4\pi}\oint_{\Sigma_\infty}D^\nu k^{\mu}d\Sigma_{\mu\nu}, \quad
J_\infty = -\frac1{8\pi}\oint_{\Sigma_\infty}D^\nu m^{\mu}d\Sigma_{\mu\nu}.
\lb{koMJ}
 \ee
We also need conserved electric charges which for the action  (\ref{eq:general_action}) follows from the modified Maxwell equations (\ref{eq:general_action}):
\be  \lb{Qu}
 Q_I^\infty=   \frac1{8\pi}\oint_{\Sigma_\infty} G^{\mu\nu}_I   d\Sigma_{\mu\nu}.
\ee

Consider some stationary asymptotically locally flat axisymmetric solution with an arbitrary rod structure $\{z_n,\, l_n\}$. Let each rod be surrounded by a thin cylinder $\Sigma_n$. Using the Gauss-Ostrogradski formula one can transform the asymptotic total mass to the sum of the local Komar masses $M_n^d$ of rods (``direct'' masses) and the bulk contribution coming from vector and scalar fields $M_\infty = \sum_n M_n^d+M_F$:
\be
    M_n^d = \dfrac1{4\pi}\oint_{\Sigma_n} D^\nu k^\mu d\Sigma_{\mu\nu},\qquad
    M_F = \frac1{4\pi}\int D_\nu D^\nu k^\mu dS_\mu, \lb{koM1}
\ee
where $\Sigma_n$ are the  timelike surfaces bounding the various sources,\footnote{We use the metric signature $(-+++)$ and the convention $d\Sigma_{\mu\nu}=1/2 \sqrt{|g|}\,\epsilon_{\mu\nu\lambda\tau} dx^\lambda dx^\tau$ with $\epsilon_{t\rho z\varphi}=1$ in Weyl coordinates. We will label $t, \varphi$ by an index $a$, and the remaining coordinates $\rho, z$ by $i,j$.  The two-dimensional Levi-Civita symbol $\epsilon_{ij}$ is defined with $\epsilon_{\rho z}=1$.} and the second integral is over the bulk.
Similarly, the asymptotic angular momentum reads
$J_\infty = \sum_n J_n^d+J_F$, where
\be
 J_n^d=- \dfrac1{8\pi}\oint_{\Sigma_n} D^\nu m^\mu d\Sigma_{\mu\nu}
,\qquad J_F=- \frac1{8\pi}\int D_\nu D^\nu m^\mu dS_\mu. \lb{koJ1}
 \ee
Using the Killing lemma for $k$,
 \be
D_\nu D^\nu k^\mu = -[D_\nu,D^\mu]k^\nu = - {R^\mu}_\nu k^\nu,
 \ee
and similarly for $m$, and an explicit form for the Ricci tensor  (\ref{Ric}),
one can
express the bulk integrals as
\begin{subequations}
\begin{align}
    \lb{MF}&
    M_F =- \frac{1}{8\pi}\int K_{IJ}\left(F^I_{it}F^{Jit}-F^I_{i\varphi}F^{Ji\varphi}\right)\sqrt{|g|}d^3x,
    \\
    \lb{JF}&
    J_F = \frac{1}{8\pi}\int K_{IJ}F^I_{i\varphi}F^{Jit}\sqrt{|g|} \,d^3x.
\end{align}
\end{subequations}
Using integration by parts, it can be represented as
\begin{subequations}
\label{eq:field_integrals}
\begin{align}
    M_F = &
    -\frac{1}{8\pi}\int \left[
        \partial_i\left(
            \sqrt{|g|}K_{IJ}\left(
                  F^{Jit} A_t^I
                - F^{Ji\varphi} A_\varphi^I
            \right)
        \right)
    \right]
    d^3x
    \\\nonumber &
    \underbrace{
        +\frac{1}{4\pi}\int \left[
              \partial_i\left(K_{IJ}\sqrt{|g|} F^{Jit} \right)A_t^I
            - \partial_i\left(K_{IJ}\sqrt{|g|}F^{Ji\varphi} \right)A_\varphi^I
        \right]
        d^3x
    }_{\mathcal{I}_M},
\\
    J_F = &
    \frac{1}{8\pi}\int
        \partial_i\left(K_{IJ}F^{Jit}A^I_{\varphi}\sqrt{|g|}\right) \,d^3x
    \underbrace{
    - \frac{1}{4\pi}\int
        \partial_i\left(K_{IJ}F^{Jit}\sqrt{|g|}\right)A^I_{\varphi} \,d^3x
    }_{\mathcal{I}_J}.
\end{align}
\end{subequations}
With the help of the modified Maxwell equations, the extracted integrals are
\begin{subequations}
\label{eq:i_integrals}
\begin{align}
    \mathcal{I}_M = &
    -\frac{1}{2\pi}\int\, d^3x\,
    \partial_i \left[\sqrt{-g} H_{IJ} \left(
        A_t^I \tilde{F}^{Jit}
        - A_\varphi^I \tilde{F}^{Ji\varphi}
    \right)\right]
    \\\nonumber& +
    \frac{1}{2\pi}\int\,d^3x\,\sqrt{-g}\, H_{IJ}
    \left(
        F_{it}^I \tilde{F}^{Jit} - F_{i\varphi}^I \tilde{F}^{Ji\varphi}
    \right),
    \\
    \mathcal{I}_J = &
    \frac{1}{2\pi}\int \,d^3x\,
    \partial_i\left(
        \sqrt{|g|} A^I_{\varphi} H_{IJ} \tilde{F}^{Jit}
    \right)
    - \frac{1}{2\pi}\int \,d^3x\,\sqrt{-g}\, H_{IJ} \tilde{F}^{Jit} F^I_{i\varphi}.
\end{align}
\end{subequations}
As far as the expressions $F^{I}_{it}{\tilde F}^{Jit}-F^{I}_{i\varphi}{\tilde F}^{Ji\varphi}$ and $F^{I}_{i\varphi}{\tilde F}^{Jit}$ symmetrized over $(I,J)$ are identically zero:
\begin{subequations}
\begin{align}
    &
    F_{it}^I \tilde{F}^{Jit} - F_{i\varphi}^I \tilde{F}^{Ji\varphi}
    =
    \frac{\epsilon^{tij\varphi}}{\sqrt{-g}} \left(
        F_{i\varphi}^J F_{jt}^I - F_{i\varphi}^I F_{jt}^J
    \right)
    \xrightarrow[]{\text{symm. }(I, J)} 0,
    \\&
    \tilde{F}^{Jit} F^I_{i\varphi} =
    \frac{\epsilon^{tij\varphi}}{2\sqrt{-g}} \left(
        F^{J}_{i\varphi} F^I_{j\varphi} - F^I_{i\varphi} F^{J}_{j\varphi}
    \right)
    \xrightarrow[]{\text{symm. }(I, J)} 0,
\end{align}
\end{subequations}
second integrals in each expression (\ref{eq:i_integrals}) become identically zero, leaving only the first integrals. Collecting all terms together, one can rewrite (\ref{MF}) and (\ref{JF}) in the full-divergence form
\begin{subequations}
\begin{align}
    \label{eq:field_contribution_m}
    & M_F =
    -\frac{1}{8\pi}\int
        \partial_i\left[
            \sqrt{-g}
            \left(
                A_t^I G_I^{it} - A^I_\varphi G_I^{i\varphi}
            \right)
        \right]
    d^3x,\\
    & J_F =
    \frac{1}{8\pi}\int
    \partial_i\left(
        \sqrt{|g|} A^I_\varphi G_I^{it}
    \right)
    \,d^3x,
\end{align}
\end{subequations}
suggesting that the bulk contributions to mass and angular momentum can be attached to surface integrals around the rods leading to the following representation of the asymptotic mass and angular momentum in terms of dressed contributions of rods:\footnote{We assume that the integrals (\ref{eq:field_integrals}) over the sphere at infinity vanish, otherwise they must be added too. This happens if the North and South Misner and Dirac strings are arranged non-symmetrically, by adding suitable constants to the asymptotic values of the rotation function $\omega$ and the azimuthal component of the four-potential $A_\varphi$.}
\begin{subequations}
\begin{align}
   & M_\infty = \sum_{n} M_n,\qquad
    M_n = \frac1{8\pi}\oint_{\Sigma_n}\,
    \left(
          g^{ij}g^{ta}\partial_jg_{ta}
        + A^I_t G_I^{it}
        - A^I_\varphi G_I^{i\varphi}
    \right)d\Sigma_i
 \\
  &  J_\infty = \sum_{n} J_n,\qquad
    J_n = -\frac1{16\pi}\oint_{\Sigma_n}\,\left(
        g^{ij}g^{ta} \partial_jg_{\varphi a}
        + 2 A^I_\varphi G_I^{it}
    \right)
    d\Sigma_i,
\end{align}
\end{subequations}
where we have also rewritten the direct Komar integrals in an explicit form. Thus, we have succeeded in presenting the bulk contributions in the form of the integrals over the rods in the same way as it was done in Ref. \cite{Clement:2019ghi} for the Einstein-Maxwell system. The scalar fields are hidden in the definitions of the induction tensors (\ref{Gdef}) and do not enter explicitly.

This procedure can be regarded as rod's dressing by the bulk fields. So the resulting Smarr formulas for the asymptotic mass and angular momentum are written as sums of the dressed rod's contributions. Note, that in the general case, this is the only representation for the global $M_\infty$, $J_\infty$ that can be found, since different forms of the rod contribution do not allow one to write asymptotic quantities directly in terms of physical charges as was possible for a single black hole.

From now on we assume the standard Weyl-Papapetrou parametrization of the metric
 \begin{equation}\lb{weyl}
ds^2 = -f(dt-\omega d\varphi)^2 + f^{-1}[e^{2k}(d\rho^2+dz^2)+
\rho^2d\varphi^2],
\end{equation}
and  the four-potential
\begin{equation}
\,A^I_\mu dx^\mu = v^I dt + a^I d\varphi.
\end{equation}
The twist potential satisfies two-dimensional dualization equation
\begin{equation}\lb{twist}
      \partial_i \chi =
    - f^2 \rho^{-1} \epsilon_{ij} \partial_j \omega
    + u_I\partial_i v^I
    - v^I \partial_i u_I,
\end{equation}
where $i, j=\rho,z$. The dualization equation for magnetic potentials reads\footnote{The normalization of the sigma-model potentials in this paper differs from these in Ref. \cite{Clement:2019ghi}.}
\begin{equation}\lb{u2}
    \partial_i u_I
    = f^{-1} e^{2k}\rho \epsilon_{ij} G_I^{j\varphi}.
\end{equation}
Computation of the ``direct'' parts of the rod Komar integrals was given in Ref. \cite{Smarr:1972kt}. Expressing them in terms of the sigma-model potentials, we obtain
\begin{subequations}
\begin{align}\lb{Md}
    & M^d_n =
    \frac1{4}\int_{z_n}^{z_{n+1}} dz \,
        \omega \left(
              \partial_z \chi
            + v^I \partial_z u_I
            - u_I\partial_z v^I
        \right),
\end{align}
\begin{align}\lb{Jd}
    &J^d_n =
    -\frac1{8}\int_{z_n}^{z_{n+1}} dz \, \omega \left\{
          2
        - \omega \left(
              \partial_z \chi
            + v^I \partial_z u_I
            - u_I\partial_z v^I
        \right)
    \right\}.
\end{align}
\end{subequations}
The electromagnetic potentials $v^I$ and $u_I$ appear in the expressions for direct rod mass and angular momentum due to the use of the dualized twist potential, i.e., implicit use of Einstein equations. These terms will be exactly canceled by the field contributions to the same quantities.

From (\ref{G0}) and the definition of the rod potentials  one finds the following relations:
\begin{subequations}
\begin{align}
    &\lim_{\rho\to 0} \left(G^{\rho t}_I-\omega G^{\rho \varphi}_I \right)\sqrt{|g|} =0, \lb{tfi}\\
    &\lim_{\rho\to 0} \left (v^I+a^I/\omega\right)\equiv - \Phi^I=\rm{const,}\lb{Ph}
\end{align}
\end{subequations}
where $\Phi^I$ is the electric potential of the rod. Together with the piecewise constancy along the coordinate $z$ of the rod angular velocity $\Omega=1/\omega$, the last relation also gives
\be \lb{Av}
\partial_z a^I=  -\omega \partial_z v^I
\ee
as $\rho\to 0$. Using Eqs. (\ref{tfi}), (\ref{Av}), and the dualization equation (\ref{u2})
one can rewrite the field contribution to the dressed rod mass   (\ref{eq:field_contribution_m}) as
\begin{align}\lb{MnF}
    & M^{F}_n
    =
    - \frac{1}{4}\int_{z_n}^{z_{n+1}} dz (
        \omega \left( v^I \partial_z u_I - u_I \partial_z v^I \right)
        - \partial_z (a^I u_I)
    ).
\end{align}
 Combining this with the direct Komar mass (\ref{Md}) we obtain the dressed rod mass
 \be\lb{Mrn}
M_n =
      \frac{1}{4}\int_{z_n}^{z_{n+1}} dz \,
    \left(
        \omega \partial_z \chi + \partial_z (a^I u_I)
    \right).
 \ee
For the bulk contribution to the angular momentum, one also applies Eq. (\ref{tfi}) and then the magnetic dualization equation (\ref{u2}). The subsequent transformations are the same as in Ref. \cite{Clement:2019ghi} and lead to the same result
\begin{align}
   J_n^F &=
      \frac{1}{4}\oint_{\Sigma_n}\,
      \left[
          \omega \partial(a^I u_I)
        + \omega^2  u_I \partial v^I
    \right]
    dz,
\end{align}
Combining with the direct Komar momentum (\ref{Jd}), we obtain the momentum of the dressed rod:
\begin{align}\lb{Jrn}
    &J_n =
    \frac{1}{8}\int_{z_n}^{z_{n+1}} dz  \partial_z \left[
        \omega \left(a^I u_I - 2 z\right)
        + \omega^2 \left( \chi - \Phi^I u_I \right)
    \right]
\end{align}
Now consider the Komar electric charge of a rod (\ref{Qu}). Explicitly, it is given by the flux
 \be\lb{Qun}
Q_{In} = \frac1{8\pi}\int_{\Sigma_n}\sqrt{|g|}G_I^{t\rho}dzd\varphi.
 \ee
 Manipulating with indices and using dualization equation (\ref{udef})
  \be\lb{indi}
G_I^{t\rho} = \frac{g^{\rho\rho}}{g_{tt}}G_{I\,t\rho} -
\frac{g_{t\varphi}}{g_{tt}}G_I^{\varphi i} = e^{-2k}\left[G_{I\,t\rho}
+ \frac{f\omega}\rho \partial_z u\right],
 \ee
one finds that in the limit $\rho\to 0$ only the second term contributes to the electric charge
\begin{align}\lb{Qn}
    Q_{In} & =
    \frac{1}{4}\int_{\Sigma_n}\omega \partial_z u_I dz.
\end{align}
Formulas for mass (\ref{Mrn}), angular momentum (\ref{Jrn}), and electric charge (\ref{Qn}) do not differ from those of the Einstein-Maxwell system and do not contain scalar terms. After integration over $z$ we get
\begin{subequations}\lb{Arn1}
\begin{align}\lb{Mrn1}
M_n =
&  \frac{\omega_n}{4} \chi\Big {\vert}^{z_{n+1}}_{z_n} +\frac{1}{4} (a^I u_I) \Big{\vert}^{z_{n+1}}_{z_n},
 \end{align}
\begin{align}\lb{Jrn1}
  J_n =
 \frac{\omega_n}{8}  \left\{
    - 2(z_{n+1}-z_n)
    + \left[
          \omega_n \left( \chi - \Phi^I_n  u_I\right)
        + a^I u_I
    \right] \Big {\vert}^{z_{n+1}}_{z_n}
 \right\},
\end{align}
\be\lb{Qrn}
    Q_{In}= \frac{1}{4} \omega_n u_I \Big {\vert}^{z_{n+1}}_{z_n}.
\ee
\end{subequations}
The angular velocity of a rod is defined as a limit
\be
    \Omega_n=\lim_{\rho\to 0}\ \omega^{-1}(\rho,z),\qquad  z_n <z<z_{n+1}.
\ee
Combining the above formulas we obtain rod Smarr mass
\be\label{leSm}
M_n=\frac{L_n}{2}+2\Omega_n J_n+\Phi^I_n Q_{In},
\ee
where $L_n=z_{n+1}-z_n$ is the $n$-s rod length (for infinite rods some length regularization is needed). Please, note that this formula is an identity; it is valid for any rod, regardless of the specific parameter values.

Although it looks like a one-dimensional object, any rod with finite surface gravity has, because $g_{\varphi\varphi} = -f\omega^2 \neq0$ for $\rho=0$, a finite cylindrical area:
\be\lb{Arn}
{\cal{A}}_n=\oint d\varphi\int_{z_n}^{z_{n+1}} \sqrt{|g_{zz}g_{\varphi\varphi}|} dz =2\pi \int_{z_n}^{z_{n+1}}\sqrt{|e^{2k|}}|\omega| dz.
\ee
The surface gravity (by definition, positive) in Weyl coordinates reads
\be\label{kappa_weyl}
\kappa_n=|\Omega_n|\sqrt{|e^{-2k}|}.
\ee
Thus, the integral (\ref{Arn}) is simply proportional to the rod length:
\be\lb{kapA}
 \frac{\kappa_n}{2\pi}{\cal{A}}_n=  {z_{n+1}-z_n}=L_n .
\ee
This is true for both timelike rods (horizons) and spacelike ones (defects). In the first case, the surface gravity is proportional to the Hawking temperature, and the area of the horizon is proportional to the entropy. For a single black hole, this gives
$
T_H = \kappa_H/2\pi,\; S_H={\cal{A}}_H/4,
$
therefore
\be\lb{TS}
T_HS_H = \frac{\kappa_H}{8\pi}{\cal{A}}_H= \frac{L_H}4,
\ee
and the Smarr formula \cite{Smarr:1972kt} can be rewritten
as
 \be\lb{smarr}
M_H = 2\Omega_H J_H + 2T_H S_H + \Phi_H^I {Q_I}_H.
 \ee
For spacelike rods thermodynamical identification does not seem justified, so we prefer to leave Smarr formula in its ``length'' form (\ref{leSm}). But still similarity between the Smarr formulas for black holes (\ref{smarr})
and strings   is remarkable:
 \be\lb{Ssmarr}
M_s = 2\Omega_s J_s + 2\kappa_s {\cal A}_s + \Phi_s^I {Q_I}_s.
 \ee

For multicenter solutions, these mass relations hold separately for each black hole constituent and each interconnecting string.
As stated in Ref. \cite{Clement:2017otx,Clement:2019ghi}, this is also true when black holes have magnetic and/or NUT charges. In our interpretation, the contribution of these parameters to the total asymptotic mass comes from individual rods representing the Dirac and Misner strings. Note that neither magnetic potentials, nor NUT-potentials enter explicitly in
Smarr-type identities, being encoded in expressions for string angular momenta and string electric potentials.

Although direct application of the above formulas is relevant for defects carrying an angular momentum associated with NUTs, it was shown  \cite{Clement:2023xvq} that a careful limit can be performed in the NUTless case to 
yield consistent mass formulas for interconnecting strings carrying only magnetic flux (Dirac strings). This extends to the present supergravity context. The string rotation parameter $\omega_S$, proportional to the difference between the NUT charges of the two adjacent black holes, goes to zero in the NUTless limit. The string angular momenta $J_S$ and string electric charges $Q^I_S$ then also vanish. In this limit, equation (\ref{elpot}) shows that the vector potentials $A^I_{\varphi S} = a^I_S$ must go to constant values 
\be
a^I_S \equiv 4P^I_S
\ee
corresponding to the magnetic fluxes exchanged between the two adjacent black holes. Equation (\ref{Mrn}) for the string mass then goes, in the limit $\omega_S\to0$, to
\be
M_S = P^I_S\Psi_{IS},
\ee
where $\Psi_{IS} = \delta_Su_I$ are the magnetic potential differences between the two ends of the string. So, contrary to the NUTty case, in the NUTless case the magnetic ``charges'' (more correctly the magnetic fluxes exchanged between neighboring black holes) and conjugate ``potentials'' (more correctly scalar potential differences) do enter the Smarr relations, but only those for the Dirac strings.


\section{Particular cases}
As aligned multi-black hole solutions are known so far only in Einstein-Maxwell theory, we restrict with one-center solutions with Misner and Dirac strings.
Most of the physically meaningful one-center supergravity solutions can be presented in the Kerr form with Boyer-Lindquist coordinates
\begin{equation} \label{eq:examples_metric}
    ds^2 =
    - \frac{\Delta - a^2 \sin^2\theta}{\Sigma} \left(dt - \omega d\varphi\right)^2
    + \Sigma \left(
        \frac{dr^2}{\Delta} + d\theta^2 + \frac{\Delta \sin^2\theta}{\Delta - a^2 \sin^2\theta} d\varphi^2
    \right),
\end{equation}
where $\Delta$ is a function of $r$, and $\Sigma$, $\omega$ functions of $r, \theta$. Their particular form as well as the relevant vector and scalar fields  depends on mass, NUT charge and electric and magnetic potentials.  To move from Boyer-Lindquist coordinates to Weyl coordinates (\ref{weyl}), one has to perform the  transformation
$\rho^2   =   \Delta   \sin^2\theta,\; z  = ( \Delta'/2) \,\cos\theta$,
and introduce the metric function
\begin{equation}
    e^{2k} =
    \frac{
        \left(\Delta - a^2 \sin^2\theta\right)^2
    }{
        \Sigma  \left(
              \Delta
            - \left(
                \Delta - {\Delta'}^2 / 4
            \right) \sin^2\theta
        \right)
    }.
\end{equation}
If $\Delta(r)$ has a root $r=r_H$ corresponding to the horizon, solving the equation $\rho=0$ gives us solutions for three rods:
\begin{itemize}
    \item  $l_+:$ northern string $r\in(r_H,\infty)$, $\cos\theta = 1$;
    \item $l_H$ the horizon rod $r = r_H$, $\theta\in[0,\pi]$;
    \item $l_+$ southern string $r\in(r_H,\infty)$, $\cos\theta = -1$.
\end{itemize}
Northern and southern strings may represent Misner and Dirac strings \cite{Clement:2019ghi}. In the case of the multi-center solutions (so far known only in  ${\cal N}=2$ supergravity), the rod structure is more complicated (see Ref. \cite{Clement:2023xvq}).

Here  we  list  some particular cases that suits the considered general model (\ref{eq:general_action}) with the corresponding coefficient matrices:
\begin{enumerate}
    \item Einstein-Maxwell (EM)
    \begin{align}
        f_{AB} = 0,\qquad
        K_{IJ} = 2,\qquad
        H_{IJ} = 0.
    \end{align}
    The Smarr formula for electrovacuum solutions is known for a long time and it was originally obtained by Smarr in Ref. \cite{Smarr:1972kt} for electrically charged rotating black holes (Kerr-Newman). The generalization to the more general rods that allows for NUT solutions was given in Ref. \cite{Clement:2019ghi}.

    \item Einstein-Maxwell-dilaton (EMD)
    \begin{align}
        f_{AB} = 4,\qquad
        K_{IJ} = 2e^{-2\alpha\phi},\qquad
        H_{IJ} = 0,
    \end{align}
    where $\phi$ is a dilaton field and $\alpha$ is a coupling constant. The black hole solutions obtained within the EMD theory with $\alpha=\sqrt{3}$ are interesting due to the existence of the polynomial constraints on the physical charges \cite{Rasheed:1995zv,Bogush:2020obx}.

    \item Einstein-Maxwell-dilaton-axion (EMDA)
    \begin{equation}
        f_{AB} = \begin{pmatrix}
            4 & 0 \\ 0 & e^{4\phi}
        \end{pmatrix},\qquad
        K_{IJ} = 2e^{-2\phi},\qquad
        H_{IJ} = \kappa,
    \end{equation}
    where $\phi$ and $\kappa$ are a dilaton and an axion scalar fields correspondingly. This model was analyzed in Ref. \cite{Galtsov:2021iqx} in details.

    \item STU-model
     \begin{align}
     &f_{AB}={\rm diag}\left( 1,\,1,\,1,\,\e^{2\varphi_1} \e^{2\varphi_2} \e^{2\varphi_3}\right),\quad H_{IJ}=-\frac12 \kappa_1
     \begin{pmatrix}
     &&&1\\&&1&\\&1&&\\1&&&
     \end{pmatrix},
     \end{align}
    \begin{equation*}
        K_{IJ} = e^{-\varphi_1}\left(
              e^{ \varphi_2 + \varphi_3}s^{(1)}_I s^{(1)}_J
            + e^{ \varphi_2 - \varphi_3}s^{(2)}_I s^{(2)}_J
            + e^{-\varphi_2 + \varphi_3}s^{(3)}_I s^{(3)}_J
            + e^{-\varphi_2 - \varphi_3}s^{(4)}_I s^{(4)}_J
        \right),
    \end{equation*}
    where
    \begin{align}
        &
        s^{(1)}_I = \left\{1, \kappa_3, \kappa_2, -\kappa_2\kappa_3\right\},
        &
        s^{(2)}_I = \left\{0, 1, 0, -\kappa_2\right\},
        \\\nonumber&
        s^{(3)}_I = \left\{0, 0, 1, -\kappa_3\right\},
        &
        s^{(4)}_I = \left\{0, 0, 0, 1\right\}.
    \end{align}
    It is worth noting that $\det (e^{\varphi_1} K_{IJ}) = 1$, and $K_{IJ}$ must be positive definite, otherwise there would be ghosts in the STU-model. This matrices follow the formulation of the STU theory symmetric under four $U(1)$ vectors $A_\mu^I,\,I=1,2,3,4$ \cite{Chow:2014cca} inherited from $S^7$ compactification of the eleven-dimensional supergravity. The theory also contains three dilatons $\varphi_a$ and three axions $\kappa_a$ combining into the six-dimensional sigma-model for the variables $\Psi_A=\varphi_a,\,\kappa_a,\,A=1,...,6$, non-linearly transforming under $SL(2,R)^3$ duality group.
\end{enumerate}

\section{Conclusion}
We generalized the derivation of Smarr mass formulas for black hole solutions containing string singularities on the polar axis to the generic bosonic supergravity action  (\ref{eq:general_action}) using   Tomimatsu's approach introduced by him within the framework of the Einstein-Maxwell theory. Crucial to this generalization is the existence of a three-dimensional sigma-model for describing stationary configurations. Allowing for string singularities, we cover solutions with Misner and Dirac strings
.

The main difference from the Einstein-Maxwell case is the presence of scalar fields.
However, it has been shown that Smarr-type relations for both black holes and strings do not contain corresponding scalar charges, and the final formulas are identical to those in the Einstein-Maxwell theory up to additional summation over electric charges. Unlike previous mass formulas for single black holes, our relations also do not contain explicit magnetic and NUT fluxes. The corresponding contributions to the global mass and angular momentum of the configuration are included in the parameters of the rods associated with the Dirac and Misner strings. In the NUTless limit, magnetic fluxes do occur explicitly, but only in the Smarr relations for the Dirac strings. 

This may seem to contradict the electric-magnetic duality of the action and proposed previously for single black hole mass formulas that share this symmetry. But, as previously shown \cite{Clement:2023xvq}, in a system of multiple aligned black holes, magnetic charges cannot be unambiguously assigned to individual black holes. Thus, multiple-center dyonic solutions spontaneously violate the electric-magnetic duality of the supergravity Lagrangian.

Our treatment also offers a different interpretation of the NUT contribution to the black hole entropy, presenting the corresponding terms in mass relations as mechanical work on stretching the Misner strings.

\begin{acknowledgments}
The work of DG was partially supported by  the Strategic Academic Leadership Program ``Priority 2030'' of the Kazan Federal University. IB and DG are grateful to Kirill Kobialko for the fruitful discussions and his advice.
\end{acknowledgments}

\appendix

\section{Sigma-model reconstruction}\label{app:sigma}
\renewcommand{\theequation}{\Alph{section}.\arabic{equation}}
For reader's convenience we sketch the derivation of the sigma-model in Refs. \cite{Breitenlohner:1987dg,Breitenlohner:1998cv} in the original variables. Starting with the four-dimensional action
\begin{equation}\label{S4}
    S_4 = \frac{1}{2} \int d^4x\sqrt{|g|} \cdot \left(
        - R_4
        + \frac{1}{2}\left<J_\mu, J^\mu\right>
        - \frac{c}{4} G_{\mu\nu}^T \mu G^{\mu\nu}
        + \frac{c}{4} G_{\mu\nu}^T \nu \tilde{G}^{\mu\nu}
    \right),
\end{equation}
where $\mu$, $\nu$ are symmetric matrices dependent on the scalar fields, and $\mu$ is positive definite. The field $G_{\mu\nu}$ is a matrix-valued Maxwell two-form (do not confuse with the induction tensor in the main part of the paper), and $J_\mu$ represents some arbitrary matrix-valued currents. The signature used in \cite{Breitenlohner:1998cv} is ${(+---)}$, which is not the same to this paper. The tilde denotes the dualization (in Ref. \cite{Breitenlohner:1998cv} it is denoted by an asterisk; there are some other deviations from the original notations, but they are insignificant and can be easily read off). Introduce an auxiliary field
\begin{equation}\label{eq:app_H}
    \tilde{H}_{\mu\nu} = \eta \left(\mu G_{\mu\nu} - \nu \tilde{G}_{\mu\nu}\right),
\end{equation}
where $\eta$ is an arbitrary constant orthogonal matrix, and $\nabla_\mu \tilde{H}^{\mu\nu} = 0$ is the modified Maxwell equations. The fields $G_{\mu\nu}$ and $H_{\mu\nu}$ can be expressed through the vector potentials: ${G_{\mu\nu} = \partial_{[\mu} B_{\nu]}}$, ${H_{\mu\nu} = \partial_{[\mu} C_{\nu]}}$ (antisymmetrization uses weight 1). Both two-forms can be combined into a block matrices:
\begin{equation}
\mathds{F}_{\mu\nu} = \begin{pmatrix} G_{\mu\nu} \\ H_{\mu\nu} \end{pmatrix},\qquad
\mathds{A}_\mu = \begin{pmatrix} B_{\mu} \\ C_{\mu} \end{pmatrix},\qquad
\mathds{F}_{\mu\nu} = \partial_{[\mu}\mathds{A}_{\nu]}.
\end{equation}
We use double-stroke symbols for the matrix representation. Since the two-form $H_{\mu\nu}$ can be expressed through $G_{\mu\nu}$ and $\tilde{G}_{\mu\nu}$ and vice versa, there is a relation $\mathds{F}_{\mu\nu} = \mathds{Y} \mathds{M} \tilde{\mathds{F}}_{\mu\nu}$, where:
\begin{equation}
    \mathds{Y} = \begin{pmatrix}
        0 & \eta^T \\
        -\eta & 0
    \end{pmatrix},\qquad
    \mathds{M} = \begin{pmatrix}
        \mu + \nu \mu^{-1} \nu & \nu \mu^{-1}\eta^T \\
        \eta \mu^{-1}\nu & \eta \mu^{-1} \eta^T
    \end{pmatrix},\qquad
\end{equation}

Consider a timelike Killing vector $k^\mu = \partial_t$. The metric and vectors can be split in the following manner
\begin{equation}
    g_{\mu\nu} = \begin{pmatrix}
        \Delta k_j k_i - \frac{h_{ij}}{\Delta}& \Delta k_j \\
        \Delta k_i & \Delta
    \end{pmatrix},\qquad
    B_{\mu} = \begin{pmatrix}
        B_i + B k_i \\
        B
    \end{pmatrix},\qquad
    C_{\mu} = \begin{pmatrix}
        C_i + C k_i \\
        C
    \end{pmatrix},
\end{equation}
where $i$ stands for the spatial directions and we omit the $t$-index:
\begin{equation}
    \mathds{A} = \begin{pmatrix}
        B \\ C
    \end{pmatrix},\qquad
    B=B_\mu k^\mu=B_t,\qquad
    C=C_\mu k^\mu=C_t.
\end{equation}
One can define a twist vector
\begin{equation}
    \tau_i = -f^2 h_{il} \frac{\epsilon^{ljk}}{\sqrt{h}} \partial_j k_k,
\end{equation}
which satisfies the following equation
\begin{equation}
    \tau_i = \partial_i \psi + \frac{c}{2} \mathds{A}^T \mathds{Y}^{-1} \partial_i \mathds{A},
\end{equation}
where $\psi$ is a twist potential. Moreover, from the definition (\ref{eq:app_H}) follows that $C$ can be interpreted as a magnetic potential
\begin{align}
    \frac{\Delta}{\sqrt{h}} \epsilon^{ijk} \partial_k C
    =
    \eta \left(\mu G^{ij} - \nu \tilde{G}^{ij}\right).
\end{align}

Then, according to Ref. \cite{Breitenlohner:1998cv}, the final three-dimensional action reads
\begin{equation}\label{eq:app_sigma_model}
    S_3 = \int d^3x\sqrt{|h|} h^{ij} \cdot \frac{1}{2} \left(
          R^3_{ij}
        - \frac{1}{2}\left<J_i, J_j\right>
        + \frac{c}{2\Delta} \partial_i \mathds{A}^T \mathds{M} \partial_j \mathds{A}
        - \frac{1}{2\Delta^2} \left(
            \partial_i \Delta \partial_j \Delta + \tau_i \tau_j
        \right)
    \right).
\end{equation}
Rewriting the quadratic form $d\mathds{A}^T \mathds{M} d\mathds{A}$ explicitly, we get:
\begin{align}
    &
    d\mathds{A}^T \mathds{M} d\mathds{A} =
    \begin{pmatrix} dB^T & dC^T \end{pmatrix}
    \begin{pmatrix}
        \mu + \nu \mu^{-1} \nu & \nu \mu^{-1}\eta^T \\
        \eta \mu^{-1}\nu & \eta \mu^{-1} \eta^T
    \end{pmatrix}
    \begin{pmatrix} dB \\ dC \end{pmatrix}
    = \\\nonumber & =
      dB^T (\mu + \nu \mu^{-1} \nu) dB
    + 2 dB^T \nu \mu^{-1}\eta^T dC
    + dC^T \eta \mu^{-1} \eta^T dC
\end{align}

Choosing $\eta = I$ and $c=2$ for simplicity, we get
\begin{subequations}
\label{eq:magnetic_twist_ama_1}
\begin{equation}
    \frac{\Delta}{\sqrt{h}} \epsilon^{ijk} \partial_k C
    =
    \mu G^{ij} - \nu \tilde{G}^{ij},\qquad
    \tau_i
    =
      \partial_i \psi
    - \left(B^T \partial_i C - C^T\partial_i B\right),
\end{equation}
\begin{align}
    d\mathds{A}^T \mathds{M} d\mathds{A} =
      dB^T (\mu + \nu \mu^{-1} \nu) dB
    + 2 dB^T \nu \mu^{-1} dC
    + dC^T \mu^{-1} dC
\end{align}
\end{subequations}

Corresppondence with our notation is as follows:
\begin{equation}
    G \to F^I,\qquad
    B \to v^I,\qquad
    C \to u_I,\qquad
    \Delta \to f,\qquad
    k_i \to -\omega_i,\qquad
    \psi \to - \chi,
\end{equation}
\begin{equation*}
    \left<J_\mu, J_\nu\right> \to f_{AB} \partial_\mu \Psi^A \partial_\nu \Psi^B,\qquad
    \mu \to K_{IJ},\qquad
    \nu \to -2H_{IJ},
\end{equation*}
and the action (\ref{S4})  with $\eta = I$ and $c=2$ transforms to  (\ref{eq:general_action}) up to common factor 1/2. Formulas in Eq. (\ref{eq:magnetic_twist_ama_1}) take the form
\begin{subequations}
\begin{equation}
    \tau_i
    =
    f^2 h_{il} \frac{\epsilon^{ljk}}{\sqrt{h}} \partial_j \omega_k
    =
    u_I\partial_i v^I - v^I \partial_i u_I
    - \partial_i \chi,
\end{equation}
\begin{align}
    \frac{f}{\sqrt{h}} \epsilon^{ijk} \partial_k u_I
    =
    K_{IJ} F^{Jij} + 2 H_{IJ} \tilde{F}^{Jij}
    =
      K_{IJ} F^{Jij}
    + H_{IJ} F^J_{\mu\nu} \frac{\epsilon^{ij\mu\nu}}{\sqrt{-g}},
\end{align}
\begin{align}
    d\mathds{A}^T \mathds{M} d\mathds{A} =
      (K_{IJ} + 4 H_{IK} K^{KL} H_{LJ}) dv^I dv^J
    - 4 H_{IK} K^{KJ} dv^I du_J
    + K^{IJ} du_I du_J.
\end{align}
\end{subequations}
Combining these expressions with (\ref{eq:app_sigma_model}), we get the sigma-model (\ref{eq:sigma_model}).

\end{document}